\documentclass[letter,traditabstract]{aa}  
\pdfoutput=1
\usepackage{graphicx}
\usepackage[caption=false,position=top,font=scriptsize,labelformat=empty,aboveskip=0pt,captionskip=0pt]{subfig}
\usepackage{amssymb}
\usepackage{float}
\usepackage{natbib}
\bibpunct{(}{)}{;}{a}{}{,}
\usepackage{txfonts}

\newcommand{\Msun} {M$_{\odot}$}

\begin{document}

   \title{Old pre-main-sequence stars}

   \subtitle{Disc reformation by Bondi-Hoyle accretion}

   \author{P. Scicluna\inst{1,2} \and G. Rosotti\inst{3,4,5} \and J.E. Dale\inst{4} \and L. Testi\inst{2,4,6} 
          }

   \institute{
ITAP, Universit\"at zu Kiel, Leibnizstr. 15, 
24118 Kiel, Germany
\and
European Southern Observatory, Karl-Schwarzschild-Str. 2,
        D-85748 Garching b. M\"unchen, Germany
\and
Max-Planck-Institut f\"ur extraterrestrische Physik,
Giessenbachstra\ss{}e, D-85748 Garching, Germany
\and
Excellence Cluster Universe, Boltzmannstr. 2, D-85748 Garching,
Germany
\and
 Universitats-Sternwarte M\"unchen, Scheinerstra\ss{}e 1,
D-81679 M\"unchen, Germany
\and
INAF-Osservatorio Astrofisico di Arcetri, Largo E. Fermi 5, I-50125 Firenze, Italy
             }
\offprints{pscicluna@astrophysik.uni-kiel.de}
   \date{Received 17 February 2014; Accepted 21 May 2014}

  \abstract{Young stars show evidence of accretion discs which evolve quickly and disperse with an e-folding time of $\sim$ 3Myr. 
   This is in striking contrast with recent observations that suggest evidence for numerous $>30$~Myr old stars with an accretion disc in large star-forming complexes. 
   We consider whether these observations of apparently old accretors could be explained by invoking Bondi-Hoyle accretion to rebuild a new disc around these stars during passage through a clumpy molecular cloud. 
  We combine a simple Monte Carlo model to explore the capture of mass by such systems with a viscous evolution model to infer the levels of accretion that would be observed. 
  We find that a significant fraction of stars may capture enough material via the Bondi-Hoyle mechanism to rebuild a disc of mass $\gtrsim$ 1 minimum-mass solar nebula, and $\lesssim 10\%$ accrete at observable levels at any given time. 
  A significant fraction of the observed old accretors may be explained with our proposed mechanism. 
  Such accretion may provide a chance for a second epoch of planet formation, and have unpredictable consequences for planetary evolution. }


   \maketitle
%

\section{Introduction\label{sec:intro}}

Circumstellar discs form around protostars as a result of angular momentum conservation
during gravitational collapse \citep[e.g.][]{1987ARA&A..25...23S}. 
In the early phases of star formation, disc material loses angular momentum and is accreted onto the central star. %
The most direct observational signature of the presence of a protoplanetary disc is the excess emission, on top of the expected naked stellar photosphere, at infrared and millimetre wavelengths, in the ultraviolet and in optical/infrared emission lines. 
The long wavelength emission is produced by a dusty disc, heated by internal dissipation processes or reprocessing of stellar radiation \citep[e.g.][]{2007prpl.conf..555D}. 
The short wavelength excess and the optical/infrared emission lines are thought to be produced by the disc-star interaction as matter accretes onto the star or is ejected in a wind/jet \citep[][]{2009apsf.book.....H}. 
Strong observational evidence shows that both the inner dusty disc and accretion onto the central star quickly disappear during the early stages of pre-main-sequence evolution; the fractions of stars with near infrared excess and with accretion signatures decay with an e-folding time of 2-3~Myr \citep{2010A&A...510A..72F,2007ApJ...662.1067H}. 
This disc dissipation timescale, even considering the possible revision by \citet{2013MNRAS.434..806B}, sets a stringent constraint on the timescales for planet formation.

Recent work has challenged this paradigm. 
Sensitive, wide field  H$\alpha$ surveys of large star-forming complexes in the Magellanic Clouds and our own Galaxy have revealed a population of pre-main-sequence stars that appear to be older than 10~Myr but still show prominent H$\alpha$ emission and/or infrared excess \citep{2010ApJ...720.1108B,2013ApJ...775...68D,2011ApJ...740...11D,2011ApJ...739...27D,2013MNRAS.435.3058D}. 
Although some of these ``old'' accretor candidates in nearby star-forming regions have been shown to be misclassified young stellar objects \citep{2013A&A...558A.114M}, it is difficult to believe that this is the case for all the candidates; these populations of old accretors are not as centrally condensed as the young stellar clusters in the same fields \citep[e.g.][]{2011ApJ...740...10D}. 
If the line emission is interpreted as due to accretion as in young pre-main-sequence stars, the implied accretion rates are similar to those derived at early ages{, and typically higher than nearby transitional discs}\footnote{{although these are systematically lower mass objects}}. 
These findings are hard to understand in a framework in which the primordial disc is still the reservoir of accreting material at such old ages; even one disc of age $>30$Myr implies an initial population $>10^5$ (assuming exponential decay with an e-folding timescale of 3Myr).

In this paper we explore the possibility that the old accretors do not have a primordial disc, but a disc that they re-accreted after the primordial disc had dissipated. 
Previous studies \citep{2009ApJ...707..268M,2005ApJ...622L..61P,2008AJ....135.2380T} have investigated the influence of Bondi-Hoyle accretion on pre-main-sequence mass-accretion rates and the protoplanetary disc at earlier phases, during the initial evolution of the disc-star system within the progenitor cloud. 
Here we investigate the possibility that a star older than 5-10Myr happens to travel through a clumpy molecular cloud, typically unrelated to that in which the star formed, and is able to accrete enough material to form a new accretion disc.

\section{Modelling\label{sec:mod}}
\subsection{Bondi-Hoyle accretion\label{sec:BH}}
\citet{1939PCPS...35..405H}, \citet{1944MNRAS.104..273B} and \citet{1952MNRAS.112..195B} proposed a mechanism by which objects can capture matter from the interstellar medium. 
A massive object moving through the ISM causes a perturbation, pulling material toward the object. 
As the capture of material is roughly symmetrical with respect to the direction of motion of the star, much of the angular momentum of the material cancels out, and hence it is captured by the star to eventually be accreted \citep{1980MNRAS.191..599D}.

The rate at which material is captured is given by
\begin{equation}
\dot{M}_{\rm BH} = \mu n \mathrm{v} \pi R^2_{\rm BH}, \label{eq:BH}
\end{equation}
\noindent where $\mathrm{v}$ is the relative velocity between the star and the ISM, $n$ is the number density of the ISM, and $\mu$ is the mean molecular weight (usually taken as $2.3 m_\mathrm{H}$).
The gravitational cross-section is given by $\pi R^2_{\rm BH}$, where $R_{\rm BH}$ is the Bondi-Hoyle radius
\begin{equation}
R_{\rm BH} = \frac{2GM_{\ast}}{\mathrm{v}^2 + c^2_{\rm s}};
\label{eq:RBH}
\end{equation}
\noindent $c_{\rm s}$ is the sound-speed of the ISM, typically $0.3\mbox{ km s}^{-1}$. For a 1\Msun\,star moving at $1\mbox{\rm km s}^{-1}$, $R_{\rm BH} \sim$ 1500 au.

\begin{table}
\caption{Parameters for Monte Carlo models}
\label{tab:MCpars}
\centering
\begin{tabular}{l l l l}
\hline\hline
Parameter & Values & Parameter & Values\\ \hline
 & \\
 $f_{\rm V}$ & $10^{-2},10^{-3},10^{-4},10^{-5}$ & $c_{\rm s}$ & $0.3 \mbox{ km s}^{-1}$\\

 $N_{\rm stars}$ & $10^5,10^5,10^6,10^7$ & $\sigma_{\mathrm{v}}$ & $1\mbox{ km s}^{-1}$\\
 $R_{\rm cl}$ & $0.1\mbox{ pc}$ &  $\alpha$ & 2.35 \\
 $n_{\rm cl}$ & $10^4 \mbox{ cm}^{-3}$ \\
 \hline

\end{tabular}
\end{table}

To explore the effect of this process in reconstituting discs around young stars, we build a simple Monte Carlo model to treat interactions between stars and clumps with densities typical for molecular clouds.
We assume a stationary clumpy molecular cloud, which we model as a collection of identical spherical clumps with radius $R_{\rm cl}$ and density $n_{\rm cl}$. 
We parametrise the density of clumps through a volume filling factor of dense gas $f_{\rm V}$. 
We assume a population of ``old'' young stars that has lost their primordial disc enters the cloud and moves through the clumpy medium. 
By randomly generating stars with masses between 0.7\Msun\,and 3.2\Msun\footnote{Stars above $\sim3$\Msun\,have strong winds which make a simple model inappropriate, while observations of old accretors are incomplete for stars below 0.7-1\Msun\, depending on the distance to the observed region.} from a Salpeter IMF \citep[$M \propto M^{-\alpha}$][]{1955ApJ...121..161S} and velocities generated assuming a velocity dispersion of $\sigma_{\mathrm{v}} = 1\mbox{ km s}^{-1}$, we sample the parameters required in Eq.~\ref{eq:RBH} from the values given in Table~\ref{tab:MCpars}.
The model simulates 10Myr treated as a series of quasi-static time steps of length $t_{\rm st} = 2R_{\rm cl} / \mathrm{v}_\ast$, assuming that each star is independent.
For each star, we calculate $R_{\rm BH}$, the volume swept out per time-step $V_{\rm st} = \mathrm{v}_\ast t_{\rm st}\times \pi \left(R_{\rm cl} + R_{\rm BH}\right)^2$, and hence the probability of encountering a dense clump \begin{equation}p = \frac{V_{\rm st} \times f_{\rm V}}{\left(4/3\right)\pi R^3_{\rm cl}}. \label{eq:pcl}\end{equation}

\noindent In each time-step a uniform random number $\zeta$ is drawn, and the star encounters a clump when $\zeta \leq p$; the impact parameter $b$ of the encounter is given by drawing a second random number $\zeta_2$ from the same generator such that $b = \left(R_{\rm cl} + R_{\rm BH}\right)\zeta^{1/2}_2$.
We then determine the accretion rate (Eq.~\ref{eq:BH}) and resolve the stellar accretion and the clump-mass depletion on a finer time-grid of 1000 sub-steps to accurately determine the accreted mass.
Interactions where $R_{\rm BH} > R_{\rm cl}$ and grazing encounters are treated correctly by taking the projected area of intersection.
By repeating this process for $>10^{5}$ stars we build up meaningful statistics about the range of possible BH accretion histories and their probabilities.
Note that each star is modelled independently, and mass accreted by a star does not influence the mass-budget available to later stars.

The accretion histories determined by this model are then passed to a viscous evolution model (Sect.~\ref{sec:VEM}) to estimate the rate at which material is accreted by the star.

Our choice of $f_{\rm V}$ is based on a reanalysis of SPH simulations of star-forming regions including feedback mechanisms presented in \citet{2012MNRAS.424..377D,2013MNRAS.430..234D} to determine the filling factor of gas at densities higher than $10^4 \mbox{ cm}^{-3}$. 
We find that for bound clouds of similar stellar mass to the regions observed by \citet{2010ApJ...720.1108B,2013MNRAS.435.3058D}, $10^{-6} < f_{\rm V} \lesssim 10^{-3}$ irrespective of whether feedback from massive stars is included.

While this provides a useful estimate of the amount of mass captured in this way, it somewhat overestimates the total as we neglect a number of physical processes.
First, we neglect the motion of the clumps and assume that $\mathrm{v}=\mathrm{v}_\ast$ in Eq.~\ref{eq:RBH}. 
Correct treatment of the relative motions would in general reduce $R_{\rm BH}$ and hence the accretion rates.
Second, stars above 2\Msun\,have significant wind and radiation pressure that will depress the accretion rate \citep{2004MNRAS.349..678E}.
Similarly, we do not include the possible influence of the X-ray photoevaporation on the accretion, which may have an analagous effect for lower mass stars.
We also ignore the possible influence of magnetic fields, which recent studies \citep[e.g.][]{2014ApJ...783...50L} have shown may reduce accretion rates by a factor of a few.
{Likewise, we neglect structure on scales smaller than a single clump; such structure is required for a disc to form, and would reduce accretion rates relative to the homogeneous clump case treated here.}
Finally, we do not include binaries. 
However, the only influence of binarity in the context of Bondi-Hoyle accretion is to increase $R_{\rm BH}$, since binaries behave as a single object of mass $M=M_1+M_2$.

\subsection{Viscous evolution modelling\label{sec:VEM}}

Due to the angular momentum of the material accreted from the clump, which may be due to a density gradient within the clump or the rotation of the clump itself, accretion cannot proceed directly onto the star \citep{1997A&A...317..793R}. 
Therefore, the formation of a thin accretion disc is expected as the result of the viscous spreading of a thin ring. 
\citet{2008AJ....135.2380T} described the ``buffer'' effect of an accretion disc, but did not directly model it. 
We assume that the material accreted from the medium circularises at a radius $r_0 = 0.1 R_\mathrm{BH}$. 
After a single impulse of accretion onto the disc, the surface density is described by $\Sigma(r)=M_0/(2\pi) \delta(r-r_0)$, where $M_0$ is the deposited mass. 
Under the influence of an effective viscosity $\nu$ that redistributes the angular momentum in the disc, the spreading ring solution \citep{lyndenbellpringle} describes the evolution in time of this initial surface density,
\begin{equation}
\Sigma(r,t) =   \frac{GM_\ast (r r_0)^{1/4}}{3\pi r^2 \nu \Omega} \exp\left[ - \left( \frac{(r_0^{1/2}-r^{1/2})^2 r}{3t \nu} \right) \right] \exp (-\lambda) I_{1/2} (\lambda), \label{eq_spreading}
\end{equation}
where $\nu$ is the kinematic viscosity of the gas, $\Omega$ the keplerian angular speed, $I_{1/2}$ the modified Bessel function of order $1/2$, $\lambda= 2r^{3/2}/( 3 (GM_\ast)^{3/2} \nu t r_0 )$, and we have specialized the expression for the $\nu \propto r$ case. From this analytical solution, it is possible to compute the mass accretion rate onto the star $\dot{M}_\mathrm{kernel}$. 
To derive the mass accretion rate history onto the star, we convolve this function with the mass accretion rate history onto the disc:
\begin{equation}
\dot{M}_\ast(t) = \int \dot{M}_\mathrm{BH} (t') \dot{M}_\mathrm{kernel} (t-t') \mathrm{d}t'. \label{eq_convol}
\end{equation}
Given a stellar mass, the loading radius, and a law for viscosity, the evolution in time is now completely determined. 
We fix the viscosity by using the well-known \citet{AlphaViscosity} prescription, $\nu = \alpha (h/r)^2 r^2 \Omega $, where $\alpha$ is the Shakura-Sunyaev parameter and $h/r$ the aspect ratio of the disc. 
We choose typical values of $\alpha=0.01$ and $h/r =0.05 (r/1 \mathrm{AU)^{1/4}}$ \citep{2011ARA&A..49..195A}. 
Operationally, we sample Eq. \ref{eq_spreading} numerically on a space and time grid. 
We integrate over space to get the mass of the disc and we numerically differentiate the result to get the mass accretion rate kernel, which can be convolved with the Bondi-Hoyle history (Sect.~\ref{sec:BH}).

\section{Results\label{sec:res}}

\begin{figure}
\resizebox{\hsize}{!}{\includegraphics[scale=0.5,clip=true,trim=1.3cm .5cm 4.cm 16.75cm]{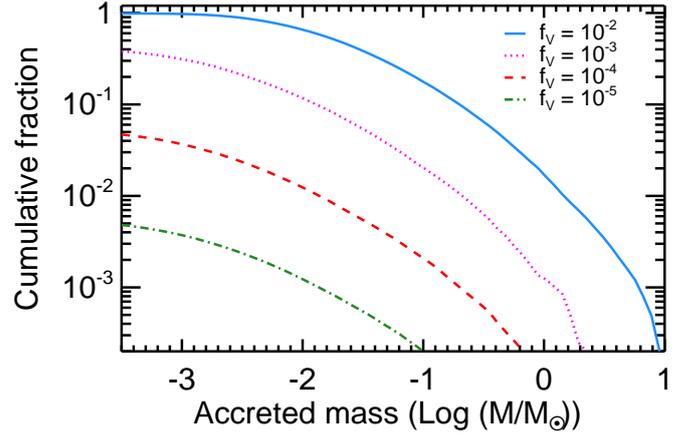}}
		\caption{Cumulative fraction of the stellar population that has accreted mass as a function of total accreted mass. The solid blue line indicates a filling-factor of $10^{-2}$, the dotted magenta line $10^{-3}$, the dashed red line $10^{-4}$, and the dot-dashed green line $10^{-5}$. } 
		\label{fig:rate}
\end{figure}

Our model indicates that a fraction of the population $\sim40-50 \times f_{\rm V}$ encounter dense regions and accrete more than $0.001$\Msun\,material by the end of the simulation (Fig.~\ref{fig:rate}). 
The median accreted mass is typically $\sim0.01$\Msun, similar to the mass of discs around young pre-main-sequence stars, with strong dependence on the stellar mass. 
In extreme cases, however, more massive stars ($>$2\Msun) with low $\mathrm{v}_\ast$ that encounter several clumps can capture $\ge$\Msun. 
Our treatment of the disc formation and evolution is probably inadequate for these extreme cases.

Converting the Bondi-Hoyle accretion into stellar accretion rates, we find $\dot{M}_{\ast} \lesssim 10^{-6}$ \Msun $\mbox{ yr}^{-1}$ after the formation of the disc. Owing to the assumptions inherent in our model, this rate declines from the peak as a power law as in primordial discs.

\begin{figure}
\resizebox{\hsize}{!}{\includegraphics[scale=0.5,clip=true,trim=1.3cm .5cm 4.cm 16.75cm]{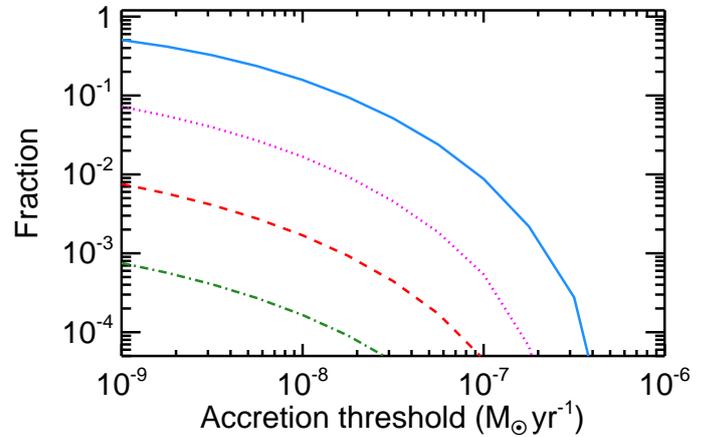}}
		\caption{Fraction of the population that would be detected as an old accretor at a given time, plotted as a function of the instantaneous accretion rate. The models are indicated using the same colours and line-styles as Fig.~\ref{fig:rate}.}
		\label{fig:thres}
\end{figure}

By calculating the time each star spends accreting above a certain threshold accretion rate, one can derive a mean time per star as a function of the threshold and hence an estimate of the fraction of the population which one expects to observe accreting at a given time.
As shown in Fig.~\ref{fig:thres}, for a threshold rate of $10^{-8}$  \Msun $\mbox{yr }^{-1}$ we typically find that the cumulative probability is $\sim 20 f_{\rm V}$, i.e. the fraction of a stellar population that one expects to observe as “old” accretors at a given time is an order of magnitude larger than the volume filling-factor of dense clumps.

\section{Discussion\label{sec:dis}}

Our primary goal is to assess whether the Bondi-Hoyle mechanism can contribute significantly to observations of old accretors in regions with ongoing star formation, under a number of simple assumptions.
This involves stars from a previous star-formation episode, after their primordial discs have dispersed, interacting with a clumpy molecular cloud.
Our model indicates that up to several percent of the population passing through a region containing dense clumps may accrete more than 0.001\Msun\,of material.
Because of the factors indicated above (Sect. \ref{sec:BH}), the model is likely to overestimate the total accreted mass.
However, since the Bondi-Hoyle accretion is a well-understood process, the largest sources of uncertainty derive from the parameters assumed as input to the model, and in particular the clump geometry and filling factor, as well as the assumption that the accreted material will form a thin disc.

Our initial choice of filling factor was based on a reanalysis of the simulations of \citet{2012MNRAS.424..377D,2013MNRAS.430..234D} for clouds similar to those observed to host old accretors.
A further estimate can be obtained from the high-resolution sub-mm maps of the 30 Dor region from \citet{2013ApJ...774...73I}. 
These reveal a wealth of clumpy structures, similar in scale and density to the clumps in the Monte Carlo model used here.
Assuming that the clumps are uniform spheres with an average radius $R_{\rm cl} = 0.15\mbox{ pc}$ and distributed in a cube whose depth is equal to the projected size of the observed region ($10 \times 10 \times 10 \mbox{ pc}^{3}$) yields a filling factor of $f_{\rm V} = 1.5\times 10^{-3}$, at the upper end of our parameter range.

The behaviour of the accretion disc depends strongly on the viscous timescale $\tau_\nu$, as parametrised in terms of $r_0$ and $\alpha$. 
An order of magnitude change in  $\tau_\nu$ has little effect on the observable old-accretor fractions at low thresholds, but the fractions at high thresholds decline approximately in proportion to $1/\tau_\nu$. 
For larger changes in viscosity, this also affects the lowest thresholds explored in Sect. \ref{sec:res}

Since we do not include stars down to the peak of IMF ($\sim$0.3\Msun) and Bondi-Hoyle accretion rates are $\propto M^{2}$, we may overestimate the total fraction of old accretors by a factor $\sim 3$ for the Salpeter IMF assumed here.
However, Eq.~\ref{eq:pcl} is dominated by $R_{\rm cl}$ for low-mass stars, so one would expect a similar fraction of old accretors when $\dot{\rm M}$ is a factor of ~4 lower. 

Comparisons between our model and the observations of old accretors are difficult, as there are no firm constraints on the size of the old population (including non-accretors). 
Nevertheless, from Fig.~\ref{fig:thres} one can see that without an unrealistically large filling factor ($\gg 10^{-3}$) of dense clumps, the small, nearby star-forming regions are unlikely to produce more than one old accretor, as their typical mass is a few hundred \Msun . 
As no old accretors have been identified in these regions, this is consistent with our model. 
From the recent identification of a large ($\sim 3\times 10^{3}$\Msun) diffuse population with ages $\gtrsim 10$ Myr toward Orion \citep{2014A&A...564A..29B} one expects a few tens of reformed discs, although it is unclear whether there is any overlap between this population and the Orion molecular clouds.

Observations of old accretors in large star-forming complexes typically detect up to several hundred such sources in each observed region.
Given the formation efficiency we have computed and our assumed filling factors, this requires a total population at least of the order of $10^{4}$ stars in the mass range of the observed old accretors, {or $\sim 3\times 10^4$ stars correcting for the IMF}, which must have passed through the regions in which the clumps are distributed.
In the case of NGC3603, which is inferred to have a population $\sim 10^{4.2}$\Msun\,\citep{2013ApJ...766..135R} {and $\sim 100$ old accretors}, this implies either that the old population was significantly richer, or that $f_{\rm V}$ is or was very high. 
The 30 Doradus region, on the other hand, shows a similar total of old accretors, although the total population is likely $\sim 100$ times larger than NGC3603.
Only a small fraction (1\%) of the stars in 30Dor need to pass through regions containing dense clumps to produce the observed numbers.
In reality, $f_{\rm V}$ will evolve with time, and it is possible that the difference we observe between these regions may be due to 30Dor being more evolved, or having evolved more rapidly, than NGC3603.

In our model, a significant fraction (up to several tens of percent) of stars capture enough material to form a circumstellar disc of mass similar to primordial protoplanetary discs. 
This raises a number of interesting questions, such as whether a second epoch of planet formation is possible, and how the interaction between inflowing material and an existing planetary system might alter the accretion or the planetary evolution.

The answers to these queries depend strongly on how the inflowing material interacts with the existing system, which we have not treated.
Nevertheless, Bondi-Hoyle accretion presents a mechanism by which a new reservoir of potentially planet-forming material may be built by up to a few percent of stars.
This gives them a second chance to form planets, from material that is potentially of different composition from the material that formed the star.
Another possibility is that these stars are already surrounded by a planetary system formed out of the primordial disc. 
If they accrete new material, typically with an angular momentum different from that of the original planetary system, the interaction of the new material and the existing planets may have a range of outcomes. 
Understanding the range of possible outcomes will require detailed simulations of the accretion process and of the dynamical interactions with the planetary systems which are beyond the scope of the present paper.


\section{Conclusions\label{sec:conc}}
We have presented a model in which Bondi-Hoyle accretion by stars passing through dense clumps in the outer regions of their natal molecular cloud leads to the re-formation of a circumstellar disc.
As a result, these stars may masquerade as pre-main-sequence objects due to ongoing accretion and the presence of infrared excess emission.
A significant part of the observed populations of old accretors in large star-forming regions may be explained by this mechanism.
As it may have wide-ranging consequences for the early evolution of planetary systems in rich stellar environments, we believe that further investigation of this mechanism is warranted.

\begin{acknowledgements}
We wish to thank the anonymous referee for their careful reading of the text.
The idea explored in this paper came up during discussions at the ESO science days and star formation coffee as well as the Munich Star Formation workshops. We thank the ESO Office for Science and all the institutes in the Munich area for providing a stimulating environment. We thank P. Armitage, G. Beccari, G. Costigan, B. Ercolano, G. De Marchi, C. Manara, N. Moeckel, A. Natta, P. Padoan, R. Siebenmorgen  and S. Wolf for discussions and insights on the various aspects discussed in this paper. 
PS is supported under DFG programme no. WO 857/10-1.
GR acknowledges the support of the International Max Planck Research School (IMPRS).
This research was supported by the DFG cluster of excellence `Origin and Structure of the Universe' (JED).
\end{acknowledgements}
\bibliographystyle{aa} 
\bibliography{BHaccretion}

\Online

\end{document}